# Optoelectronic and thermoelectric properties of Ba$_3$*D*N (*D* = Sb, Bi): A DFT investigation

Enamul Haque, Md. Taslimur Rahman, and M. Anwar Hossain

Department of Physics, Mawlana Bhashani Science and Technology University

Santosh, Tangail-1902, Bangladesh

Email: enamul.phy15@yahoo.com

**Abstract**

We have investigated the optoelectronic and thermoelectric properties of hexagonal antiperovskites Ba$_3$*D*N (*D* = Sb, Bi) using DFT calculations. The calculated equilibrium lattice parameters of both compounds are in good agreement with the available data. The calculated electronic structures indicate that they are direct bandgap semiconductors and the values of bandgaps are 1.35 and 1.33 eV for Ba$_3$SbN and Ba$_3$BiN, respectively. The inclusion of the spin-orbit effect split the conduction bands and the band gap of Ba$_3$BiN is much reduced. These two compounds have a high absorption coefficient, notably higher than that for GaAs and close to that for silicon. The obtained static refractive index is ~2.8 and 3.26, for Ba$_3$SbN and Ba$_3$BiN, respectively. We predict that both materials are suitable for a high-efficiency solar cell. Both compounds exhibit a high Seebeck coefficient and high power factor. Our analysis predicts that the studied materials are potential candidates in thermoelectric device applications.

## 1. Introduction

The energy crisis is one of the biggest challenges for 21$^{st}$ century's mankind and hence, researchers are trying to find the alternative energy sources and energy-saving materials. The solar energy is called green energy that can help the human being to face such a great challenge. Recently, perovskite materials have attracted much than other type of materials because of their suitability practical applications and safety for environment. These type of materials have a high absorbance, high tunable bandgap in the visible range [1]. Like perovskite, antiperovskite materials may also fulfill these criteria. The thermoelectric materials become a much attractive alternating source of energy although the efficiency of energy conversion by a thermoelectric power generator is rather poor and limited below 20% that of the Carnot efficiency [2]. High-performance thermoelectric materials are required in thermoelectric generators (TEG) to increase its conversion efficiency [3–5]. The performance of materials can be characterized by the dimensionless figure of merit [6], $ZT = \frac{S^2 \sigma T}{\kappa_e + \kappa_l}$ [7], where $S$, $\sigma$, $\kappa_e$, and $\kappa_l$ are the Seebeck coefficient, electrical conductivity, electronic and lattice thermal conductivity, respectively. Usually, a good thermoelectric material possesses a high thermopower and low thermal conductivity with a $ZT$ about unity or greater than unity [2,8]. The antiperovskite materials, possess a good thermoelectric performance and may suitable for practical applications [9–11]. Antiperovskite, $Ca_3SnO$ (bandgap 0.32 eV) has been experimentally found to exhibit high thermopower, approaching about ~100 μV/ K at the room temperature [9]. Recently, it has been predicted from the theoretical study that $SbNCa_3$ and $BiNCa_3$ have a figure of merit ($ZT$) around 1 at 300 K [12].

The ternary nitrides, $SbNBa_3$ and $BiNBa_3$ were reported from an experimental study that both of them adopt the hexagonal anti-perovskite type structure (P63/mmc, # 194) [13]. Recently, some authors reported the structural and optoelectronic properties of cubic $Ba_3SbN$ and $Ba_3BiN$ from the first-principles study. Another fact that they considered either GGA-PBE function or local density approximation (LDA) functional [13–17]. It is well known that GGA-PBE/LDA underestimates the experimental value of bandgap by 30-100% [18,19]. For example, the reported value of bandgap of cubic $Ba_3BiN$ by using LDA of Perdew-Wang is 0.5 eV [16], while a more accurate GW method produces a bandgap of 1.79 eV. A few other authors reported electronic properties of these compounds in the hexagonal phase by using LDA [13,14]. Since both optical and thermoelectric properties are closely related to the bandgap of the material, a precise electronic structure calculation is required to reveal the potential applications of them in different technological devices.

Here, we report precise electronic structure calculations based on DFT to study optical and thermoelectric properties by using PBE functional and different versions of BJ potential [20]. When we consider spin-orbit coupling (SOC) effect, Sb-p/Bi-p states are split and the bandgap of both compounds is reduced. The values of direct bandgap (without SOC) are 1.35 and 1.33 eV for $Ba_3SbN$ and $Ba_3BiN$, respectively. We obtain the value of static refractive index ~2.8 and 3.26 for $Ba_3SbN$ and $Ba_3BiN$, respectively. Both compounds have a high power factor and they have a great potential in thermoelectric device applications.

**2. Computational methods**

We have performed structural relaxations (minimizing forces up to 1 mRyd/au and lattice parameters simultaneously) and optoelectronic calculations within density functional theory (DFT) [21,22] by using full-potential linearized augmented plane wave method implemented in WIEN2k [23]. We have modeled muffin tin sphere of Ba, Sb/Bi and N with radius 2.5, 2.5, and 2.38 Bohr, respectively. We have performed the whole Brillouin zone (BZ) integration using $4 \times 4 \times 4$ and $21 \times 21 \times 21$ non-shifted k-point mesh for relaxations and electronic structure calculations, respectively. The exchange-correlation potential has been treated within generalized gradient approximation of Perdew-Burke-Ernzerhof (GGA-PBE) [24,25] and to overcome underestimation of bandgap by PBE, different versions of Becke-Johnson (BJ) potential have been used [19,20,26,27]. Since Sb and Bi are heavy elements, we have performed a separate electronic structure calculation including spin-orbit coupling effect for Sb/Bi. For optical and transport properties calculations, we have generated the eigenvalues using a finer $44 \times 44 \times 44$ non-shifted k-point mesh and performed optical properties calculations in WIEN2k [28]. The generated eigenvalues file has been fed into BoltzTrap code, a semi-classical Boltzmann transport equation solver within constant relaxation time approximation (RTA).

## 3. Results and Discussions

Firstly, we have relaxed the structure and calculated the ground state lattice parameters. Our calculated lattice parameters are listed in Table-1, with experimental and available theoretical values.

Table-1: Comparison of calculated lattice parameters with experimental and available theoretical values.

| Compounds | a (Å) | | | c (Å) | | |
|---|---|---|---|---|---|---|
| | Present | Exp.[a] | Other theoretical[b] | Present | Exp.[a] | Other theoretical[b] |
| $Ba_3SbN$ | 7.5672 | 7.5336 | 7.5095 | 6.6552 | 6.6431 | 6.6603 |
| $Ba_3BiN$ | 7.6667 | 7.6111 | 7.6006 | 6.7025 | 6.6791 | 6.7066 |

a: Ref. [13]
b: Ref. [14]

From the above table, it is clear that our calculated values of lattice parameters fairly agrees with the experimental values, as well as other available theoretical values. We have used these optimized structural parameters in all the subsequent calculations.

### 3.1. Electronic properties

The electronic structure is closely related to optical and transport properties. For example, a high-performance photovoltaic material should have a bandgap of 1.5 eV. Therefore, it is reasonable to study the electronic structure of $Ba_3DN$. PBE functional underestimates the experimental bandgap by 50% [18], thus, we have calculated the electronic structure using TB-mBJ potential. Since Sb and Bi are heavier atoms, we have also calculated electronic structure by including spin-orbit coupling (SOC) effect. Our calculated electron dispersion relations of both compounds along high-symmetry directions without and with spin-orbit coupling effect are shown in Fig. 1. We see that the energy bands around Fermi level of both compounds are dispersive and the valence band maxima (VBM) and the conduction band minima (CBM) are at Γ point. Therefore, both of the studied compounds are direct bandgap semiconductors. Our calculated bandgap values without SOC are 1.35 and 1.33 eV for $Ba_3SbN$ and $Ba_3BiN$, respectively. Note that the band structures of

Ba$_3$SbN and Ba$_3$BiN are almost the same because the outer shell electrons are the same for Sb and Bi. The slight difference between them arises from the energy difference between Sb-5p and Bi-6p orbitals.

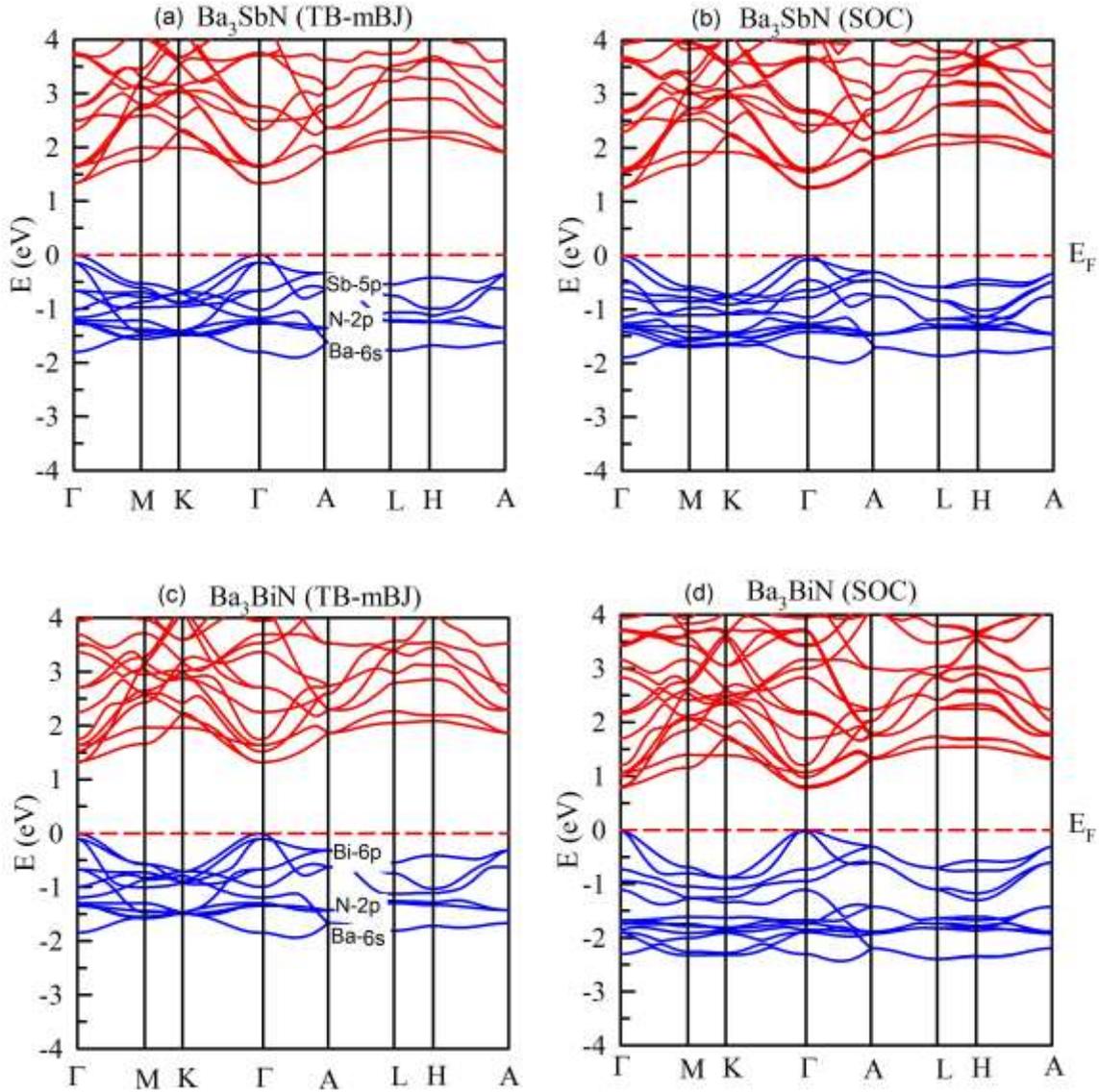

Fig. 1. Electrons dispersion relations of Ba$_3$SbN (top panel) and Ba$_3$BiN (bottom panel) with and without spin-orbit coupling (SOC) effect. The horizontal line at zero energy indicates the Fermi level.

Note that the inclusion of spin-orbit coupling effect split the conduction band, valence bands too. This type of splitting of energy bands can be visualized from the projected density of states as shown in Fig. 2 (c) and Fig. 2 (g) for $Ba_3SbN$ and $Ba_3BiN$, respectively. We see that SOC splits the p-states into $p_{1/2}$ and $p_{3/2}$ states. Thus, the energy gap between CBM and VBM is reduced and the value of it is 1.28 eV for $Ba_3SbN$ and 0.8 eV for $Ba_3BiN$. It is clear that SOC has a significant effect on the electronic structure of Bi-based compound while it is much less effective for the Sb-based compound. Because the atomic mass of Bi is much higher than that of Sb.

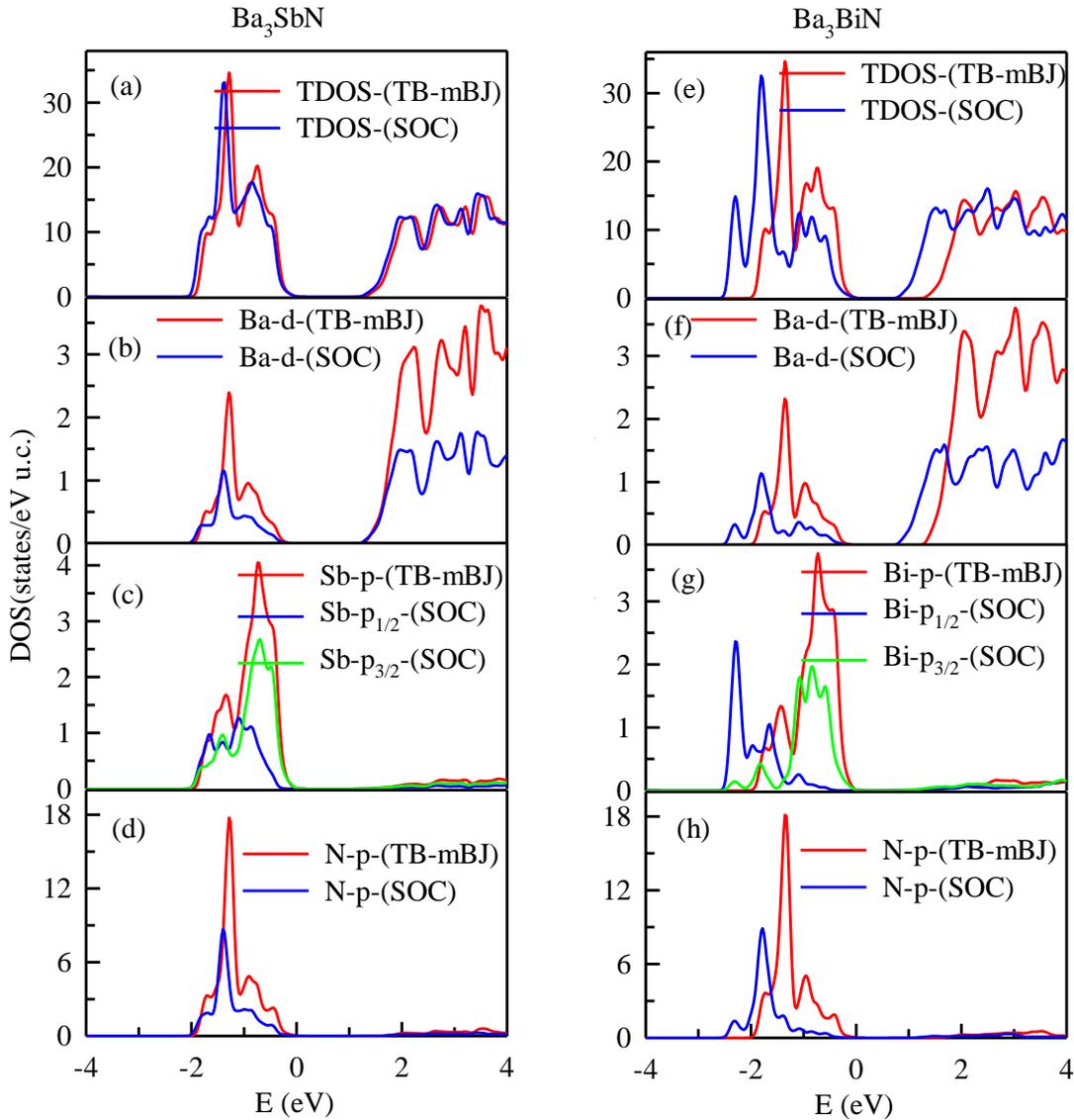

Fig. 3. Total and projected density of states (DOS): (left panel) $Ba_3SbN$ and (right panel) $Ba_3BiN$.

From the calculated total and projected density of states, we see that valence bands mainly arises from a strong hybridization between Sb-5p/Bi-6p and N-2p, and a weak hybridization between Ba-5d and strongly hybridized orbitals. The conduction bands of both compounds originate from a strong hybridization between Ba-5d and Sb-5p/Bi-6p states. Thus, Ba-5d and N-2p orbitals mainly induce direct bandgap, with a small contribution from Bi-5p. The SOC effect enforces the Ba-5d and N-2p hybridization to be weaker and Sb-5p/Bi-6 and N-2p hybridization to be stronger. The bandgap formulation in Bi-based compounds is drastically changed by SOC effect and Bi-6p states have a significant contribution in the bandgap formulation. Therefore, a heavier element like bismuth, it is essential to consider the spin-orbit coupling effect to explain electronic structure accurately. Since the bandgap of both compounds is close to the optimum value (1.5) for a material used in the photovoltaic devices, we hope that both the compounds will possess good optical performance.

### 3.2. Optical properties

For a material to be used in solar cell, it must have a high absorbance, high refractive index and low emissivity of light. Moreover, a good optical material has a high dielectric constant, high optical conductivity and reflectivity. A material with energy bandgap ~1.5 eV usually fulfills these criteria. Since our calculated bandgap of both materials is close to the optimum value, it is interesting to describe their optical properties. The interaction of light (photons) with electrons of a material can be described from the dielectric function ($\varepsilon$) of the material. Both intraband and interband transitions contribute to the dielectric function. However, intraband transitions have very small contributions to the dielectric function for a semiconductor. Thus, we have not considered in our present calculations. The dielectric function consists of two parts, related to the expression $\varepsilon(\omega) = \varepsilon_1(\omega) + i\varepsilon_2(\omega)$, namely ($\varepsilon_1(\omega)$) real and ($\varepsilon_2(\omega)$) the imaginary part. In the present case, we will only describe the dielectric function, absorption coefficient, optical conductivity, and the refractive index of both compounds for the phonon energy range 0-8 eV, with and without considering spin-orbit coupling effect.

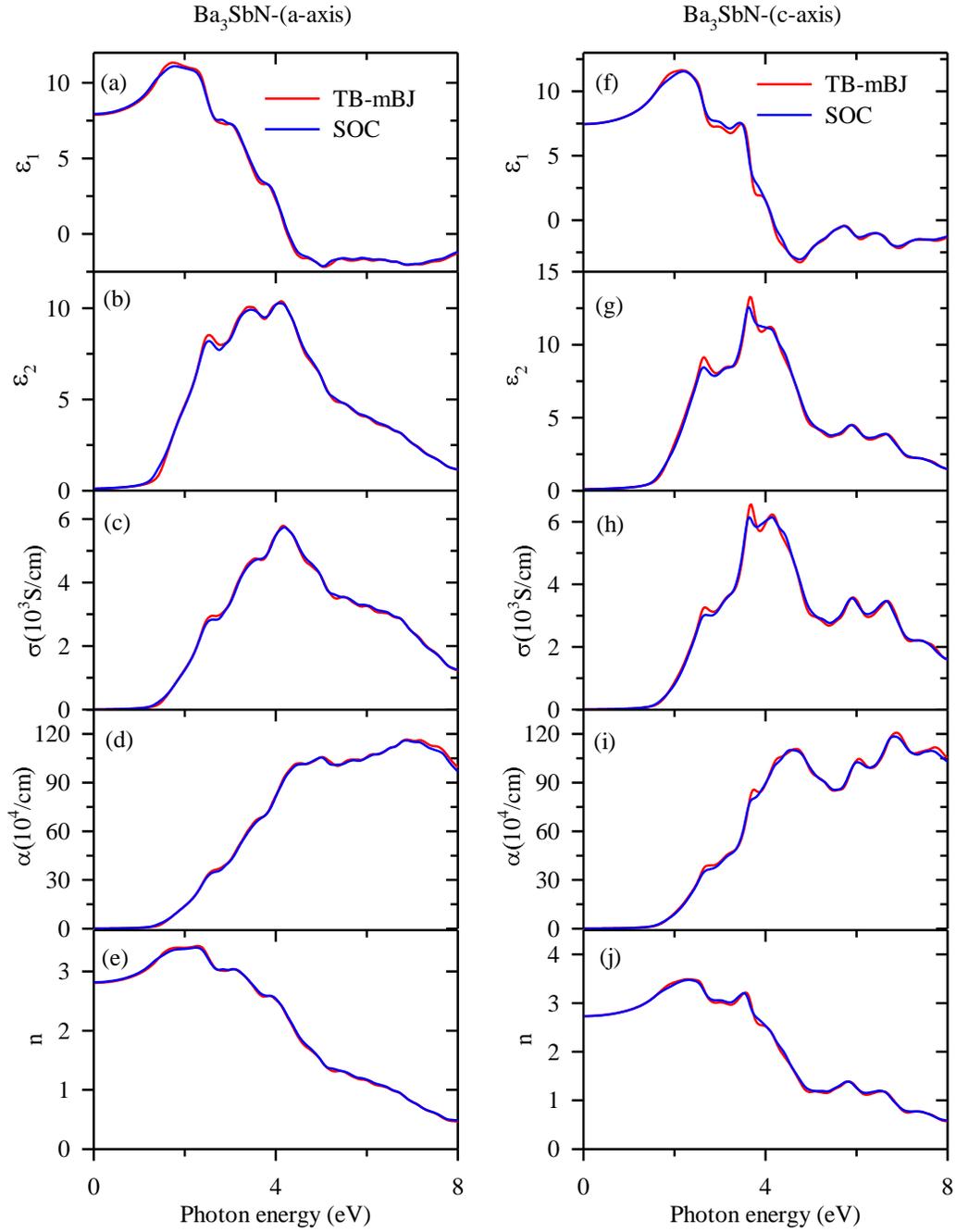

Fig. 4. The optical parameters: (a) real dielectric function $\varepsilon_1(\omega)$, (b) Imaginary dielectric function $\varepsilon_2(\omega)$, (c) optical conductivity ($\sigma$), (d) absorption coefficient ($\alpha$), (e) refractive index (n) for $Ba_3SbN$.

The calculated real and imaginary part of the dielectric function for parallel (along a-axis) and perpendicular (along-c-axis) of $Ba_3SbN$ and $Ba_3BiN$ are presented in Fig. 4-5(a, b and f, g), respectively. The similar dielectric function of $Ba_3SbN$ and $Ba_3BiN$ results from their similar type of electronic structure. We see that dielectric functions with and without considering spin-orbit coupling effect are almost same for $Ba_3SbN$ while dielectric functions with and without SOC much differ for $Ba_3BiN$. Such results are expected because electronic structure of $Ba_3BiN$ is much affected by SOC as compared to $Ba_3SbN$. It is also clear that dielectric functions show almost isotropic behavior, as the lattice parameters of both compounds along with a- and c-axis are almost the same. The major peak around 4 eV is arisen by the interband transition from Sb-5p/Bi-6p valence state to a Ba-4d conduction state. Due to the SOC effect in $Ba_3BiN$, this peak is slightly shifted to lower photon energy. This is attributed to the SOC splitting of the valence band, i.e., Bi-6p into $Bi-6p_{1/2}$ and $Bi-6p_{3/2}$. Other two small peaks around the major peak are attributed to the transitions from N-2p, the excited electrons of Ba-6s, to a Ba-5d conduction state. The calculated static real dielectric function $\varepsilon_1(0)$ are 7.89 and 7.46 for $Ba_3SbN$ along parallel and perpendicular directions, 8.5 and 7.91 for $Ba_3BiN$, respectively. These values are smaller than the values 11.7 and 12.9 for well know photovoltaic materials silicon and GaAs, respectively. The critical point of the imaginary dielectric function located around 1.35 eV describes the fundamental absorption edge of both compounds. This value of fundamental absorption edge is in a good agreement with our calculated energy bandgap of both materials.

The real part of the optical conductivities of $Ba_3SbN$ and $Ba_3BiN$ are presented in Fig. 4(c, h) and Fig. 5(c, h) for parallel and perpendicular components, respectively. Both compounds do not conduct within the photon energy range 0-1.35 eV, because of their energy bandgap. Above this photon energy range, electrons are excited to the conduction band, i.e., the interband transition between Sb-5p/Bi-6p and Ba-5d conduction band and excited electrons from Ba-6s to Ba-4d, both materials conduct and optical conductivity increases up to around ~4 eV. Above the photon energy of interband transition, the optical conductivity of both materials decreases with photon energy. The spin-orbit coupling has no effect on the optical conductivity of $Ba_3SbN$ but it improved the optical conductivity of $Ba_3BiN$ below fundamental interband transition. This is expected because SOC effect reduced the bandgap of $Ba_3BiN$ more than that for $Ba_3SbN$. The calculated maximum value of the real part of optical conductivity both compounds is around $6 \times 10^3$ S/cm at ~4 eV photon energy.

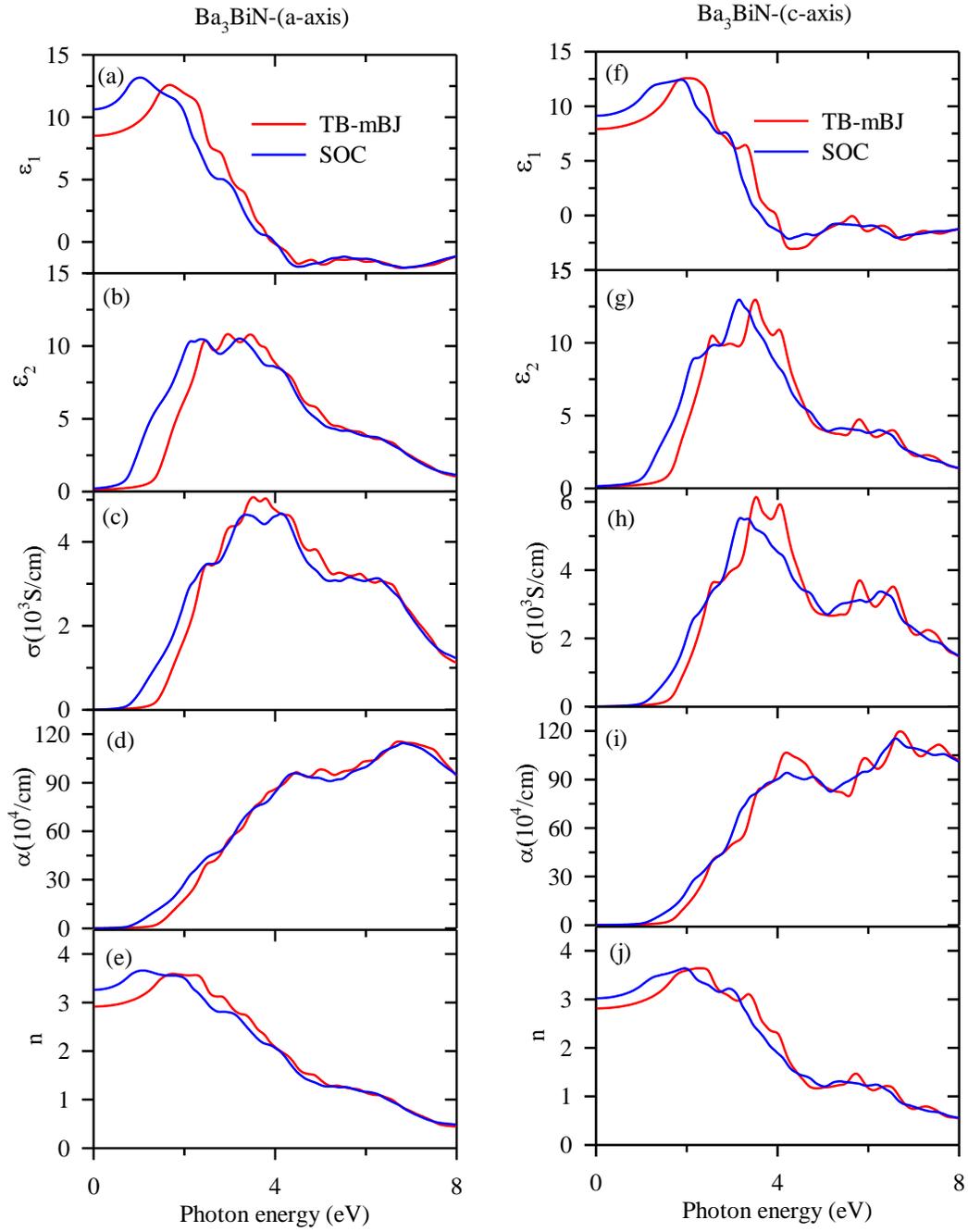

Fig. 5. The optical parameters: (a) real dielectric function $\varepsilon_1(\omega)$, (b) Imaginary dielectric function $\varepsilon_2(\omega)$, (c) optical conductivity ($\sigma$), (d) absorption coefficient ($\alpha$), (e) refractive index (n) for $Ba_3BiN$

The absorption coefficient of a material is a very important parameter that mostly affects the solar cell efficiency. Our calculated values parallel and perpendicular components of the absorption coefficient of $Ba_3SbN$ and $Ba_3BiN$ are shown in Fig. 4(d, i) and Fig. 5(d, i), respectively. We see that both compounds have no absorbance within the range of energy bandgap and absorption coefficient increases with the photon energy. Both the $Ba_3SbN$ and $Ba_3BiN$ have almost the same absorption coefficient that reflects the same electronic structure of the studied compounds. The maximum value of the absorption coefficient of the studied compounds is around $120 \times 10^4$ cm$^{-1}$ at ~8 eV for both components. This value of absorption coefficient is much larger than that of GaAs ($(14-22) \times 10^4$ cm$^{-1}$ at ~4.8 eV) [29,30] but close to the value for silicon ($1.8 \times 10^6$ cm$^{-1}$ [31]. This suggests that both compounds are promising materials for photovoltaic device applications.

The frequency dependent refractive index of parallel and perpendicular components of both compounds are presented in Fig. 4 (e, j) and Fig. 5(e, j), respectively. The refractive index of $Ba_3SbN$ shows almost isotropic behavior and the static refractive index of it is around 2.8. It increases up to ~2 eV and sharply decreases with photon energy above 2 eV. This value is much smaller than that for GaAs (3.29–3.857) [32,33] and silicon (3.88) [34]. However, the static refractive index of $Ba_3BiN$ is around 3.26 which is close to the value of GaAs (3.29–3.857) [32,33] and silicon (3.88) [34]. Therefore, $Ba_3BiN$ is a better material for photovoltaic material than $Ba_3SbN$. Our above analysis suggests that both materials are suitable for practical photovoltaic device applications.

### 3.3. Thermoelectric properties

For thermoelectric device applications, a material must have a high Seebeck coefficient, electrical conductivity, and low thermal conductivity. Both electrons and phonons contribute to the thermal conductivity. BoltzTraP code, a solver of semi-classical Boltzmann transport equation, can calculate the Seebeck coefficient, electrical conductivity and electronic thermal conductivity within the constant relation time. It calculates these parameters as a function of chemical potential, carrier concentration, and temperature. In the present paper, we will only focus on these parameters as a function of carrier density at constant temperature, 350, 500, and 700 K. However, the calculation of phonons contribution to the thermal conductivity (lattice thermal conductivity) requires the harmonic (second order) and anharmonic (third order) interatomic force calculations that are much expansive.

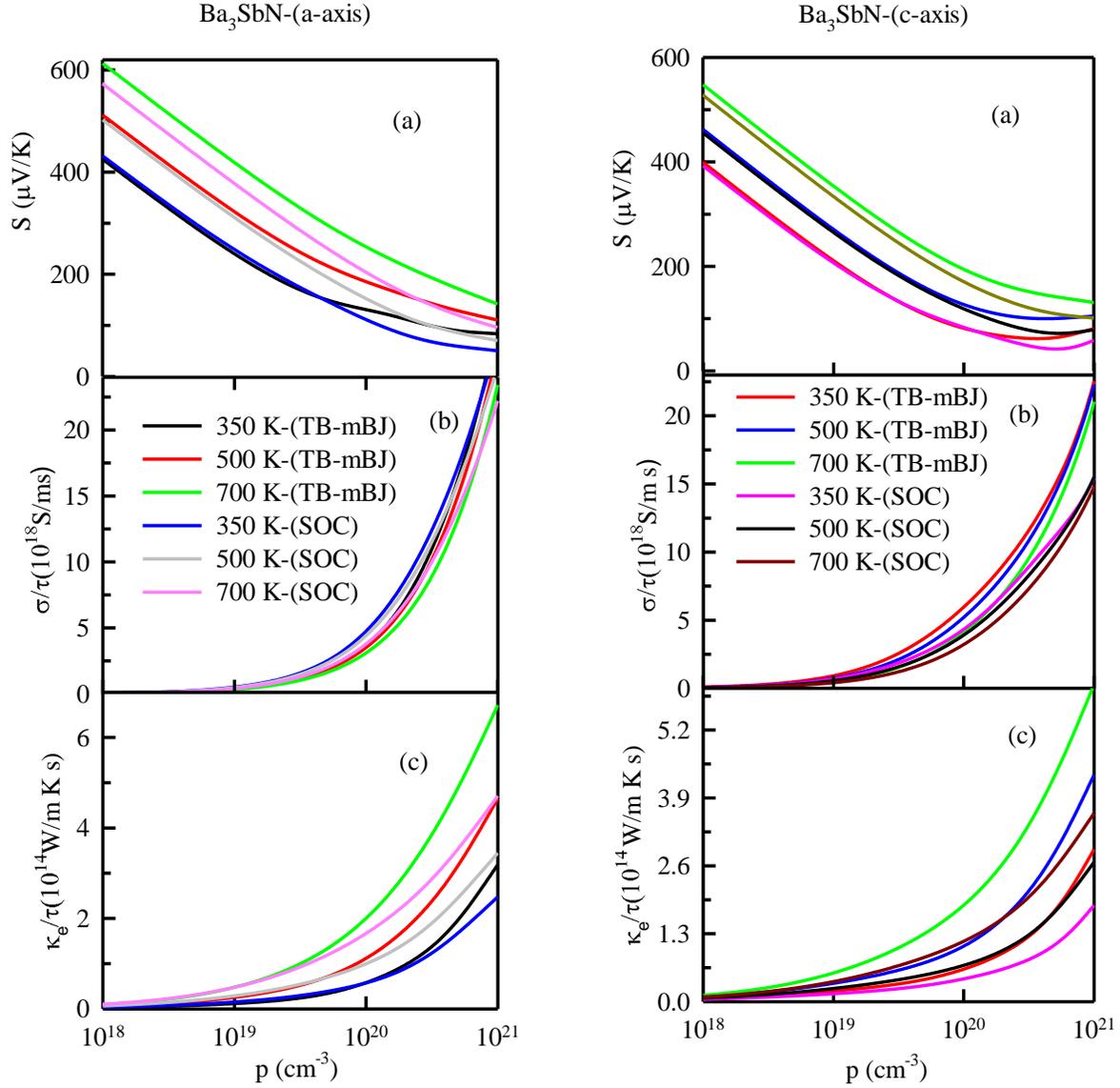

Fig. 6. Calculated (a) Seebeck coefficient (S), (b) electrical conductivity ($\sigma$), and (c) electronic thermal conductivity ($\kappa_e$) of $Ba_3SbN$, as a function of carrier density. The left panel shows parallel components (along a-axis) and the right panel shows perpendicular components of these quantities (along c-axis).

We will not describe lattice thermal conductivity and figure of merit (ZT) since ZT strongly depends on lattice thermal conductivity. Moreover, we will consider these parameters only for p-type carriers (holes, since the Seebeck coefficient is larger for holes than that for electrons, in the present case). From the above points, we have calculated the Seebeck coefficient (S), electrical conductivity ($\sigma$), electronic thermal conductivity ($\kappa_e$), and power factor ($S^2\sigma$) $Ba_3SbN$ and $Ba_3BiN$ to realize the material performance and its practical applicability. Our calculated Seebeck coefficient (S), electrical conductivity ($\sigma$), electronic thermal conductivity ($\kappa_e$) of $Ba_3SbN$ as a function of carrier density at constant temperature 350, 500, and 700 K are shown in Fig. 6. The left panel shows the parallel components (along a-axis) and the right panel shows perpendicular components (along c-axis) of these parameters. The calculated Seebeck coefficient in both directions decreases with carrier density as shown in Fig. 6 (a) and 6 (b) because of the Seebeck coefficient $S \sim m^*/n^{2/3}$. It increases with temperature since the temperature shifts the Fermi level to the middle of bandgap and increases the intrinsic activation energy. Although the spin-orbit coupling (SOC) slightly affects the electronic structure of $Ba_3SbN$ at ambient condition, we note that at a high temperature, the Seebeck coefficient is highly affected by it (SOC). The calculated Seebeck coefficient at 700 K without SOC is much larger that of with SOC. They differ much as compared to that at 350 or 500 K. The calculated Seebeck coefficient at 700 K and $10^{18}$ cm$^{-3}$ carrier density is around 600 µV/K for both components. Since the thermoelectric performance of a material is directly proportional to the square of the Seebeck coefficient, $Ba_3SbN$ may be a promising material for thermoelectric device applications. From the comparison of values of transport parameters for parallel and perpendicular components, it is clear that transport properties of $Ba_3SbN$ are almost isotropic. Such behavior is expected because the lattice parameters for both components are almost the same. The electrical conductivity increases with carrier concentration (as $\sigma = ne\mu$). But we see that electrical conductivity decreases with temperature when we simultaneously change the carrier density and temperature. The increase of carrier density along with temperature reduces the mobility of carrier and enhances the lattice scattering with carriers. It is interesting to note that parallel component of the electrical conductivity is increased due to the inclusion of spin-orbit coupling effect while the perpendicular component is reduced, as shown in Fig. 6 (b-left panel) and Fig. 6(b-right panel). Since the bandgap of $Ba_3SbN$ is reduced by SOC effect, the conductivity should be increased, as it is true for the parallel component. However, it is very hard to realize the fact why electrical conductivity along c-axis is reduced by SOC effect since we are unable to study the anisotropic electronic structure of $Ba_3SbN$.

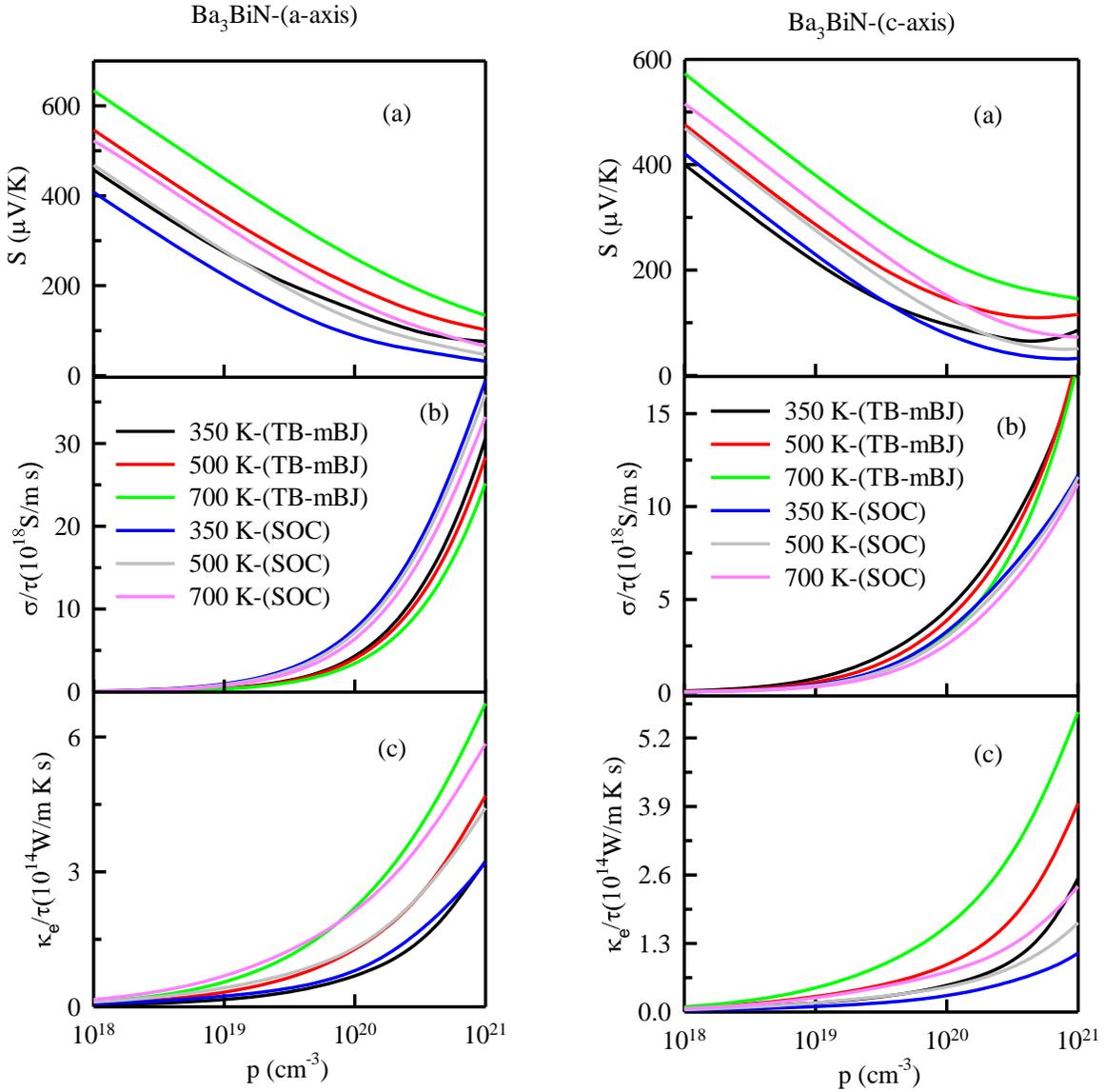

Fig. 7. Calculated (a) Seebeck coefficient (S), (b) electrical conductivity ($\sigma$), and (c) electronic thermal conductivity ($\kappa_e$) of $Ba_3BiN$, as a function of carrier density. The left panel shows parallel components (along a-axis) and the right panel shows perpendicular components of these quantities (along c-axis).

The calculated electronic thermal conductivity of Ba$_3$SbN along both directions increases with carrier density, like electrical conductivity, as shown in Fig. 6 (c-left panel) and Fig. 6(c-right panel). Like electrical conductivity, electronic thermal conductivity along both axes is almost the same. Therefore, the thermoelectric performance along a-axis may be higher than c-axis (since the Seebeck coefficient is larger along a-axis).

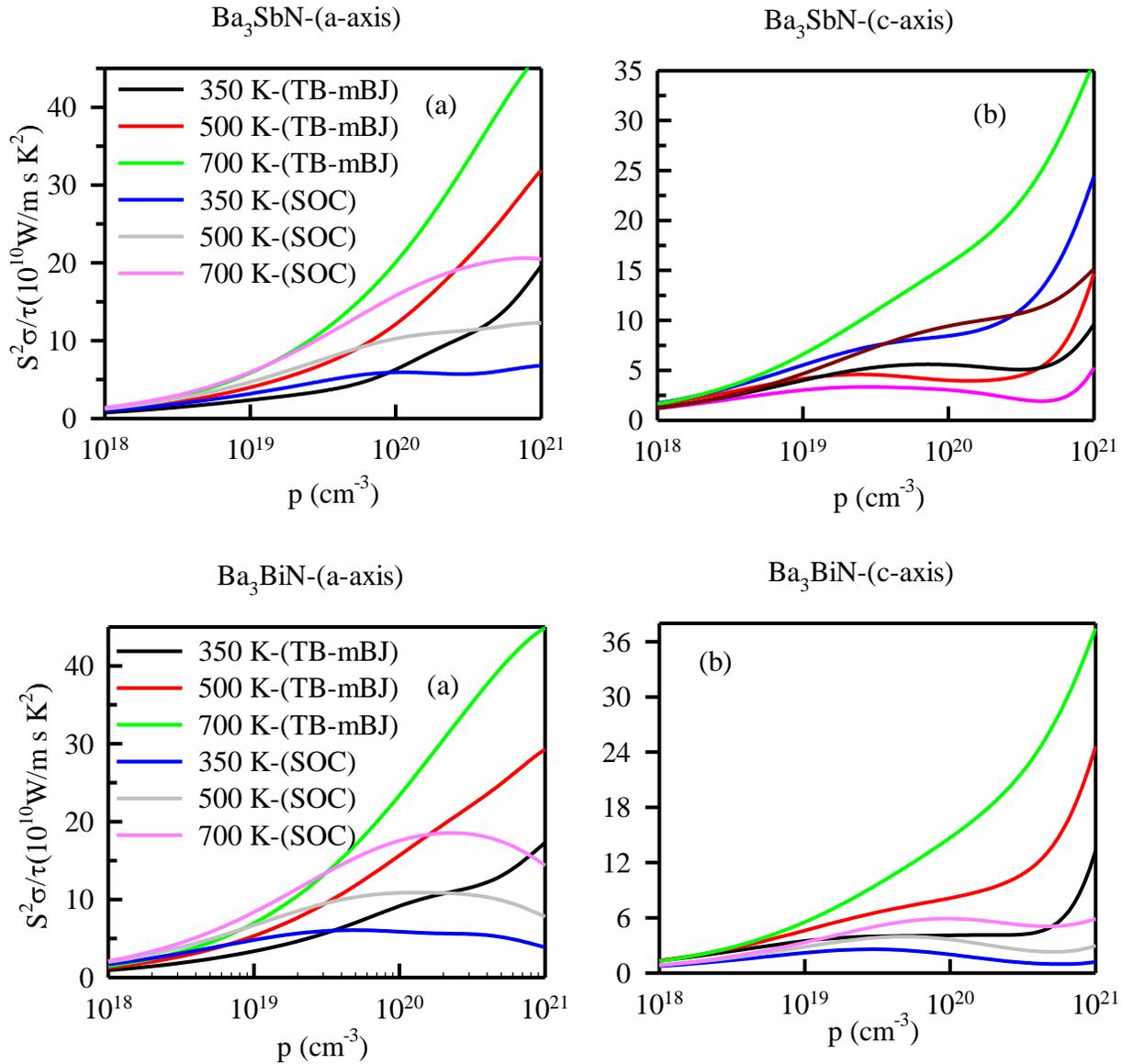

Fig. 8. Calculated power factor ($S^2\sigma$) as a function of carrier density of Ba$_3$SbN (top panel) and Ba$_3$BiN (bottom panel).

Our calculated Seebeck coefficient, electrical conductivity, and electronic thermal conductivity of Ba$_3$BiN for parallel and perpendicular components are presented in Fig. 7. Like Ba$_3$SbN, Seebeck coefficient decreases with carrier density and the Seebeck coefficient is slightly larger than Ba$_3$SbN. SOC effect reduces the Seebeck coefficient much as compared to Ba$_3$SbN since the bandgap is much reduced by SOC in Ba$_3$BiN. The calculated electrical conductivity and electronic thermal conductivity show the same trend, like Ba$_3$SbN. Although electrical conductivity along a-axis is almost two times larger than c-axis, the electronic thermal conductivity for both components is almost the same. Thus, it will result in high thermoelectric performance along a-axis than c-axis. Moreover, Seebeck coefficient of Ba$_3$BiN is slightly larger than Ba$_3$SbN. Therefore, hexagonal antiperovskite Ba$_3$BiN may be more effective thermoelectric material than Ba$_3$SbN. This can be further cleared from the calculated power factor (PF), as shown in Fig. 8. We see that power factor increases with carrier density, but the scenario is totally changed after $10^{20}$ cm$^{-3}$ in the case of Ba$_3$BiN (SOC effect). Below $10^{20}$ cm$^{-3}$, the power factor of Ba$_3$BiN is larger than Ba$_3$SbN. Above this carrier density, the Seebeck coefficient of Ba$_3$BiN is sharply dropped more than that of Ba$_3$SbN. Thus, the power factor above this carrier density of Ba$_3$SbN is slightly larger than Ba$_3$BiN. From the above analysis, we predict that both materials have a good potential in thermoelectric device applications and Ba$_3$BiN is more effective thermoelectric material than Ba$_3$SbN.

### 4. Conclusions

In summary, we have predicted optoelectronic and thermoelectric properties of hexagonal antiperovskites Ba$_3$DN (D = Sb, N) within the density functional theory. The calculated equilibrium lattice parameters of both compounds are in a good agreement with the experimental and available theoretical data. Both the compounds are direct bandgap semiconductors and the values are 1.35 and 1.33 eV for Ba$_3$SbN and Ba$_3$BiN, respectively. Due to the spin-orbit effect, the conduction bands split and the band gap of Ba$_3$SbN is slightly reduced while the band gap of Ba$_3$BiN is much reduced. Since the bandgap of both compounds is close to the optimum value (1.5 eV) that required for maximum efficient solar cells, we obtain a high value of the static refractive index (also high absorption coefficient), ~2.8 and 3.26, for Ba$_3$SbN and Ba$_3$BiN, respectively. Both the compounds possess high Seebeck coefficient and large power factor. We hope that solar cells made of them may have a high efficiency and thermoelectric performance can be further improved by alloying or nano-structuring.